%% file: main.tex
\documentclass{nature}
\usepackage{graphicx}
\usepackage{amsmath}
\usepackage{amssymb}
\usepackage{color}
\usepackage{url}
\usepackage{verbatim}
\usepackage{tabu}
\usepackage{sansmath}
\usepackage{upgreek}
\usepackage{epsfig}
\usepackage{caption}
\usepackage{url}
\usepackage[dvipsnames]{xcolor}

\DeclareCaptionLabelFormat{adja-page}{\hrulefill\\#1 #2 \emph{(previous page)}}

\def\lapp{\ifmmode\stackrel{<}{_{\sim}}\else$\stackrel{<}{_{\sim}}$\fi}
\def\gapp{\ifmmode\stackrel{>}{_{\sim}}\else$\stackrel{>}{_{\sim}}$\fi}

\newcommand{\yy}{$YY^*$}
\newcommand{\mr}{\mathrm}

\newcommand{\postsub}{}
\newcommand{\psnote}{}
\newcommand{\ef}{}
\newcommand{\vk}{}
\newcommand{\kwm}{}
\newcommand{\spt}{}

\newcommand{\bca}{}
\newcommand{\dww}{}
\newcommand{\revMD}{}
\newcommand{\revRS}{}

\newcommand{\psvtwo}{}
\newcommand{\vktwo}{}
\newcommand{\psvthree}{}
\newcommand{\sptvthree}{}

\newcommand{\src}{SGR~1935+2154}

\newcommand{\rthree}{FRB~180916.J0158+65}

\newcommand{\fitburstDM}{332.7206}
\newcommand{\fitburstDMerror}{0.0009}

\newcommand{\fitburstBurstSep}{28.91}
\newcommand{\fitburstBurstSeperror}{0.02}
\newcommand{\fitburstScattering}{0.759}
\newcommand{\fitburstScatteringerr}{0.008}
 
\newcommand{\fitburstRefFreqScat}{600} 

\title{A bright millisecond-duration radio burst \psvtwo{from a Galactic magnetar}}

\input{auth-sgr1935}

\begin{document}

\maketitle

\blfootnote{
* Corresponding Author\\
\affils
}

\begin{abstract}
Magnetars are highly magnetized young neutron stars \vktwo{that} occasionally produce enormous bursts and flares \vk{of X-rays and gamma-rays}\cite{kb17}. Of the approximately thirty magnetars currently known in our Galaxy and Magellanic Clouds,
five have exhibited transient radio pulsations\cite{ok14,erb+20}.
Fast radio bursts (FRBs) are millisecond-duration bursts of radio waves arriving from cosmological distances \cite{phl19}. Some have been seen to repeat \cite{ssh+16a,abb+19c,fab+20}.  A leading model for repeating FRBs is that they are extragalactic magnetars, powered by their intense magnetic fields \cite{lyu14,bel17,mms19}. However, a challenge to this model has been that FRBs must have radio luminosities many orders of magnitude \vk{larger} than those seen from known Galactic magnetars.  Here we report the detection 
of an extremely intense radio burst from the Galactic magnetar \src\ using the Canadian Hydrogen Intensity Mapping Experiment (CHIME) FRB project.  The fluence of this two-component bright radio burst and the estimated distance to \src\ together imply a \psvthree{400--800\,MHz burst energy of $\sim 3 \times 10^{34}$~erg}, which is \vktwo{three} orders of magnitude brighter than those of any radio-emitting magnetar detected thus far. Such a burst coming from a nearby galaxy 
would be indistinguishable from \vktwo{a typical} FRB.  This event thus bridges a large fraction of the \vk{radio \vktwo{energy} gap} between \vk{the population of Galactic} magnetars and FRBs, strongly supporting \psvthree{the notion that} magnetars are the origin of at least some FRBs.
\end{abstract}

The \vk{Canadian Hydrogen Intensity Mapping Experiment (CHIME)} radio telescope at the Dominion Radio Astrophysical Observatory in Penticton, British Columbia consists of four fixed reflecting cylinders, each 20-m by 100-m, oriented North-South, with 256 equispaced antennas sensitive to 400--800 MHz radiation. CHIME is a transit instrument with a $\sim3^{\circ}\times120^{\circ}$ instantaneous field of view. Digitized and amplified antenna signals are sent to a powerful correlator \vktwo{that} provides 1024 independent sky beams \psnote{spanning the CHIME field of view,} which \psvthree{are searched} for FRBs in real time\cite{abb+18}.

On 28 April 2020, the CHIME/FRB instrument detected a 
dispersed radio burst during a period of unusually intense X-ray burst activity \cite{pal20} from the known Galactic magnetar \src \cite{ier+16}. The burst was detected simultaneously in 93 of the 1024 CHIME/FRB formed beams, indicating an extremely bright event. The detected event \psvthree{(Figure \ref{fig:waterfall})} consisted of two sub-bursts with \psnote{best-fit temporal widths of $0.585\pm0.014$\,ms and $0.335\pm0.007$\,ms (after correcting for propagation \ef{and beam-attenuation} effects) separated by} $\fitburstBurstSep\pm\fitburstBurstSeperror$\,ms. \psnote{The best-fit estimates of the burst properties are shown in Table \ref{tab:properties}}
\psvthree{(see Methods for a description of the fitting procedure).} 
Both components show clear evidence of multi-path scattering, with a thin-screen scattering timescale of $\fitburstScattering\pm\fitburstScatteringerr$\,ms (referenced to \fitburstRefFreqScat{}\,MHz).

The detected ``comb-like'' spectral structure, as seen in Figure~\ref{fig:waterfall}, is characteristic of a \vk{CHIME} far sidelobe event, well outside the $\sim$3$^{\circ}$-wide overhead main-lobe of the telescope's primary beam.  Using an algorithm that combines the differing \psvthree{detected spectra from the} many beams, we reconstructed the burst's sky position to be \vk{(J2000)} 
RA (deg) = 293.9, Dec (deg) = +22.1, \psnote{with systematic uncertainties of order $1^\circ$ (see Methods)}, which is 0.3$^{\circ}$ from the known position of \src \cite{ier+16}, and 22$^{\circ}$ \psvtwo{west of} the CHIME meridian. This and the known ongoing intense activity from this magnetar \cite{pal20} \psnote{identify the origin \vktwo{of the burst} to be \src.}

The two burst components, \vktwo{fit jointly}, have a dispersion measure (DM) of  $\fitburstDM\pm\fitburstDMerror$\,pc\,cm$^{-3}$, determined \psvtwo{from our best-fit spectro-temporal model (Table \ref{tab:properties})}.
The maximum DM predicted from the Milky Way along this line
of sight is 500--700~pc~cm$^{-3}$, depending on the assumed Galactic electron density distribution model\cite{cl02,ymw17}. The source of the burst is thus clearly within our Galaxy, consistent with an association with \src.
The measured DM is also consistent with the source's predicted DM of $530\pm200$\,pc\,cm$^{-3}$ based on a relation between the X-ray absorbing column ($N_\mathrm{H}$) and DM\cite{hnk13} and the measured $N_\mathrm{H}$ toward \src\cite{ier+16}. Additionally, the \vktwo{Faraday} rotation measure (RM) determined for the associated supernova remnant, G57.2+0.8, suggests\cite{ksgr18} a DM of $\sim$290~pc~cm$^{-3}$, \vk{albeit with large uncertainty}.  The measured DM of the burst sits squarely among these various estimates, further supporting the association.

\begin{table} 
\caption{\textbf{Properties of burst from SGR 1935+2154.}}
\begin{center}
\begin{tabular}{l|c|c}
\hline\hline
Parameter & Component 1 & Component 2 \\
\hline
Dispersion measure (pc cm$^{-3}$) & \multicolumn{2}{c}{332.7206(9)} \\
Scattering timescale (ms)$^a$ & \multicolumn{2}{c}{0.759(8)} \\
Arrival time (UTC, topocentric)$^b$ & 14:34:24.40858(2) & 14:34:24.43755(2) \\
Arrival time (UTC, geocentric)$^{b,c}$ & 14:34:\postsub{24.42650}(2) & 14:34:\postsub{24.45547}(2) \\
Scattering-corrected width (ms) & 0.585(14) & 0.335(7) \\
Spectral index$^{a,d}$  & $-$5.75(11) & 3.61(8) \\
Spectral running$^d$ & 1.0(3) & $-$19.9(3) \\
Fluence (kJy ms) & 480 & 220 \\
Peak flux density (kJy) & 110 & 150 \\\hline
\end{tabular}
\end{center}
Values in parentheses denote statistical uncertainties corresponding to the 68.3\% confidence interval \vktwo{in the last digit(s)}.\\
$^a$ Quantities are referenced to 600 MHz. \\
$^b$ Listed arrival times were corrected for the frequency-dependent time delay from interstellar dispersion using the listed dispersion measure, and are referenced to infinite frequency. \\
$^c$ Arrival times at the geocenter were obtained after correcting the listed topocentric times for the geometric delay, assuming an ICRS source position of (R.~A., Dec.) = (19$^h$34$^m$55.606$^s$, 21$^\circ$53$^\prime$47.4$^{\prime\prime}$)\cite{ier+16}, and an observatory position of (Long., Lat.,\sptvthree{Height})$_{\rm CHIME}$ = (119$^\circ$36$^\prime$26$^{\prime\prime}$ W, 49$^\circ$19$^\prime$16$^{\prime\prime}$ N,  \sptvthree{545\,m.}).\\
\vktwo{$^d$ Quantity defined in Methods.}

\label{tab:properties}
\end{table}
\vktwo{Immediately following the CHIME/FRB detection of \src, the
10-m radio dish in Algonquin Provincial Park, Ontario, which is outfitted with a CHIME feed and which continually records baseband onto a disk buffer, was triggered to save its buffered data.  The 10-m dish is presently a testbed for real-time very-long-baseline interferometry of FRBs with CHIME.}
Analysis of these data (see Methods) provides a measurement of the RM of the event \vk{of $116 \pm 2 \pm 5$~rad~m$^{-2}$}
(measurement and systematic uncertainties, respectively)
approximately consistent with previously measured values in this direction\cite{ksgr18} as well as to the RM recently reported for a much fainter radio burst detected on 30 April 2020 by the FAST telescope\cite{zjm+20}.  
\psvthree{We find that the position angle of the linear polarization vector of the two burst components is the same to within measurement uncertainties}, and set an upper limit on any change of $<30^{\circ}$ (see Methods).

The occurrence of the burst in a far sidelobe of CHIME, where \vktwo{the sensitivity of the telescope} \revMD{has rapid spatial variation and} is not easily calibrated, makes measuring the burst's \vk{flux and fluence} challenging.
Nevertheless, using data from transits of both the Sun and the Crab Nebula (which transits 0.12$^{\circ}$ in declination \psvthree{away} from \src,
facilitating comparisons) we have estimated the frequency-dependent sensitivity of CHIME at the location of the detected burst (see Methods). From those measurements, we determine a 400--800 MHz average fluence of 480~kJy\,ms for the first burst component and 220~kJy\,ms for the second, for a combined fluence of 700~kJy\,ms.  The band-average peak flux density was 110~kJy for the first component and 150~kJy for the second. All \psvtwo{flux density and fluence measurements} are subject to \revMD{roughly} a factor 2 systematic uncertainty 
(see Methods).
Our fluence measurement is lower than the \vktwo{preliminary} $>1.5$~MJy\,ms reported at 1.4\,GHz by the STARE2 instrument\cite{bkr+20}. 
\psvtwo{Our observed spectrum of the second component rises steeply towards the top of the 400--800\,MHz band and so a higher fluence at 1.4\,GHz would be consistent if the emission is from that component.}
\vk{The distance to \src\ has been estimated\cite{ksgr18,zzc+20} to be
in the range 6.6--12.5~kpc and we assume a fiducial
10~kpc for what follows.}
\psnote{Given our measured \vktwo{peak} flux densities and this fiducial distance, the 400--800-MHz peak \vktwo{spectral} luminosity of the burst is $190^{+190}_{-90} d_\mathrm{10\,kpc}^2$ \,MJy\,kpc$^2$, assuming isotropic emission. \vktwo{The burst fluence} corresponds to an energy emitted in the 400--800\,MHz band of $3^{+3}_{-1.6}\,d_\mathrm{10\,kpc}^2\times10^{34}$\,erg and a peak \vktwo{400--800-MHz} luminosity of $7^{+7}_{-4}\,d_\mathrm{10\,kpc}^2\times10^{36}$\,erg\,s$^{-1}$.}

Several high-energy telescopes reported the detection of a \psvtwo{two-component} hard X-ray/soft gamma-ray burst at the time of the radio event from \src\cite{msg+20,rga+20,zxl+20}. \psvtwo{The reported arrival times from the {\em Insight}-HXMT telescope for the two components of the X-ray burst corrected to the geocentre are} 14:34:24.4289 and 14:34:24.4589 UTC\cite{zxl+20,zlz+20}. The two X-ray components \psvtwo{each} occur within \postsub{$\sim3$\,ms} of the \psvtwo{respective} dispersion-corrected geocentric arrival times of the CHIME burst components (see Table \ref{tab:properties}). The fluence of the soft gamma-ray burst in the 20--200\,keV band, as reported by the Konus-{\em Wind} experiment\cite{rga+20}, was $7.63(0.75)\times10^{-7}$\,erg\,cm$^{-2}$. Our measured fluence, $700^{+700}_{-350}$\,kJy\,ms, gives a radio-to-gamma-ray fluence ratio of $9^{+9}_{-5}\times 10^{11}$\,Jy\,ms\,erg$^{-1}$\,cm$^2$ (or $4^{+4}_{-1.8} \times 10^{-6}$ in \vk{dimensionless units} \psvtwo{factoring in the 400\,MHz CHIME bandwidth}). This is five orders of magnitude above the upper limit placed\cite{tkp16} on radio emission at the time of a giant flare from Galactic magnetar SGR~1806$-$20. 

The \src\ radio burst was detected during an extended active phase of the magnetar, in which hundreds of high-energy bursts were reported.  We note four other reported\cite{gcn26163,gcn26169,gcn26171} 
soft gamma-ray bursts occurred at times when the source was at a similar or smaller hour angle \psvtwo{than the detected radio burst} from the CHIME meridian, on 4 and 5 November 2019.  \vk{The two \vktwo{events} on 5 November had high-energy fluxes 14 and 11 times smaller, respectively, than that} of the 28 April 2020 event (see Methods). Yet, CHIME/FRB detected no significant radio events at those epochs, \vk{with conservative 3-$\sigma$ flux density upper limits in our band of $< 1.2$ and $< 3.0$~kJy, respectively,
implying radio to high-energy flux ratios at least 9 and 4 times smaller, respectively} (see Methods). Radio counterparts to high-energy bursts from magnetars are thus either not emitted at every burst, or extend to low flux ratios.  Alternatively, geometric effects such as beaming of the radio emission may hinder radio burst observability.

The \src\ radio burst is by far the most radio-luminous such event detected from any Galactic magnetar.
Five \vktwo{other} Galactic magnetars have been observed to emit radio pulsations\cite{ok14,erb+20}.
These pulsations 
are made up of short millisecond duration `spiky' subpulses \cite{pmp+18,erb+20}, which sometimes show spectral variations \cite{pmp+18} reminiscent of those seen in some FRBs \cite{hss+19,abb+19c,fab+20}. These, however, have thus far been observed to be many orders of magnitude fainter than the radio burst from \src.
\vk{The brightest radio burst previously seen from a magnetar (see Figure~\ref{fig:burstenergies}) was three orders of magnitude fainter; it had fluence $>200$~Jy~ms at 6\,GHz \cite{bei+18} and occurred during the 2009 outburst\cite{sk11} of magnetar 1E~1547.\sptvthree{0}$-$5408 (which showed an X-ray burst ``forest'' like that\cite{pal20} of \src). 
Thus, the April 28 event from \src\ clearly signals that magnetars can produce far brighter radio bursts than has been previously known.}

The
\src\ radio burst has implications for magnetars as potential sources of FRBs.
In Figure \ref{fig:burstenergies} we show the fluences, distances, and implied \psvtwo{burst energies} from extragalactic FRBs alongside Galactic sources of short duration radio emission, such as magnetars and pulsars.
\psvthree{Here and in Figure \ref{fig:burstenergies}, a fiducial emitting bandwidth of 500\,MHz is assumed when translating fluences to burst energies.}
Notably, the bursts from the nearest extragalactic FRB source with a well-determined distance, \rthree\ (observed fluences spanning 0.2\,Jy\,ms
\cite{mnh+20} to 37\,Jy\,ms\cite{aab+20} and luminosity distance of 149\,Mpc\cite{mnh+20}), have implied radio \psvthree{energies of $3\times10^{36}$\,erg to $4\times10^{38}$\,erg.}
Other FRBs with only DM-estimated distances, \vk{such as} FRB~181030.J1054+73\cite{abb+19c} and FRB~141113\cite{pab+18}, are potentially even less \psvtwo{energetic}, with DM-implied distances that put their \psvthree{energies as low as $\sim10^{35}$\,erg\, s$^{-1}$.}
This places the event from \src\ only 1--2 orders of magnitude below the observed \psvtwo{burst energies} of the \sptvthree{typical} FRBs \sptvthree{but the burst could have similar energies to bursts from FRB 181030 and FRB 141113 if they were at their nearest possible distance.}. As the detection of these faint FRBs is limited by the sensitivity of our instruments, the intrinsic luminosity distribution of the faintest FRB sources may overlap with that of the event from \src. This detection has therefore substantially narrowed the vast luminosity gap between what had been observed from \psnote{Galactic sources} in the past and what is observed from extragalactic FRBs. 

However, FRBs at cosmological distances can be much more energetic. FRB~180110, possibly the most luminous FRB reported to date, is at a luminosity distance of $\sim3-5$\,Gpc given its DM excess\cite{plm+19}, and was detected with a fluence of $\sim390$\,Jy\,ms\cite{smb+18}. 
This implies \psvthree{an emitted energy of $10^{42} - 10^{43}$\,erg. This is eight to nine} orders of magnitude more energetic than the burst from \src. It is as yet unclear whether such energetic events could be generated by conventional magnetars, though the total energies are plausibly within the range of magnetar energetics\cite{hbs+05}. 

The morphology of the burst from \src\ resembles those detected from
FRBs. The $\sim$1-ms \psnote{measured} durations of the subcomponents are typical of the widths of bursts from the 18 CHIME-discovered repeating FRB sources \cite{abb+19c,fab+20}. The broad spectra of the two \src\ components are strikingly different between the two components, with the first detected primarily at frequencies under $\sim$600~MHz and the second above. Though the frequency-dependent telescope response is complicated for this off-axis detection, it does not vary on the timescale of the separation \vktwo{between} the two bursts.  Extreme burst-to-burst spectral variations are common in repeating FRBs\cite{ssh+16a,fab+20}.
However, a common feature in repeating FRB morphologies is multiple subcomponents 
marching downward in frequency\cite{hss+19,abb+19c,fab+20} \sptvthree{---} opposite to what is seen in
the 28 April event from \src, where the two components increase in frequency (see Figure \ref{fig:waterfall}). The 29-ms separation between the two components is larger than \vk{is typically observed\cite{abb+19c,fab+20},}
which could signal two distinct events, rather than one with drifting.  Also, upward frequency drifting in FRBs is not completely unprecedented; \vk{for example,} one of the bursts from \rthree\ on MJD~58720 shows two components separated by 60\,ms with the second detected at higher frequencies than the first\cite{aab+20}. 

Models for magnetars as sources of FRBs fall in two main classes.
For models in which the radio waves are produced in the
magnetosphere of an active magnetar\cite{lyu02,klb17}
the short durations and
separation of the two radio bursts are natural, as is the $<$ few ms radio to gamma-ray burst time separation.
\vk{Though a magnetospheric magnetar model might suggest periodicities at the neutron-star rotation rate, the apparent absence thereof\cite{zgf+18} in one prolific FRB (121102) could be consistent} with the lack of spin-phase
dependence of bursts in some magnetars \cite{sk11}.
The other main class of magnetar model involves the neutron star as a central
engine in a surrounding nebula of past outflow material, wherein a high-energy \psvtwo{flare} creates an FRB
via a synchrotron maser blast wave at typically large ($r \simeq 10^{13}-10^{15}$ cm) distance
from the magnetar \cite{lyu14,bel17,mms19}. In such models, the high-energy emission \psvtwo{from the flare} should precede
the radio burst by the light travel time to the maser emission region, much longer than \vk{the $<1$-ms} coincidence observed
for SGR 1935+2154, unless the flow is highly relativistic and/or the dimensions are greatly scaled down in the \src\ system \vk{such that $r/\Gamma^2 \simeq 3 \times 10^7$~cm,} where $\Gamma$ is the Lorentz factor.  
The radio to high-energy energy efficiency of $\sim 10^{-6}$ is as predicted in these FRB models.
\psvtwo{For both models,} the absence of radio bursts at all high-energy burst epochs, as reported here and elsewhere\cite{tkp16}, could be due to the required relativistic beaming. \psvtwo{Downward frequency drifts\cite{lyu20}, as seen in many repeating FRB bursts\cite{hss+19,abb+19c,fab+20}, are involved in both sets of models but are unseen in this and other Galactic magnetar radio bursts.} 

\spt{Can the population of FRBs, which has burst rate\cite{bkb+18, rav19} $\approx 2\times10^3 - 2\times10^5 \,\mathrm{Gpc^{-3}\,yr^{-1}}$, be explained by a single population of \src-like magnetars?  Based on the CHIME/FRB detection of the 28 April burst from \src\ and the CHIME/FRB non-detection of bursts from nearby star-forming galaxies, in spite of significant exposure to them (see Methods), we constrain the rate of magnetar bursts with energy $E>10^{34}\,$ergs to be $0.007-0.4\,\mathrm{yr^{-1}}$ for a typical Galactic magnetar. 
The rate of bursts at $E>10^{36}\,$ergs is constrained to be $<0.004\,\mathrm{yr^{-1}}$ from the non-detection of FRBs from the Virgo cluster\cite{alf+19} (see Methods), albeit at a different radio frequency.} 

\spt{Based on these two constraints, and assuming that magnetar burst energies follow a power-law distribution with rate $R(E>E_0)\propto E_0^{\alpha}$, we calculate an upper limit to the rate of magnetar bursts with typical FRB energies ($E\gtrsim10^{38}\,$ergs) to be $<0.002\,\mathrm{yr^{-1}}$, per magnetar, \sptvthree{with a power law index limited to $\alpha\lesssim -0.15$.}}

\spt{Given the ubiquity of magnetars, which form at $\sim$10--50\% of the core-collapse supernova rate\cite{tcd+14}, or $(1-5)\times10^4\,\mathrm{Gpc^{-3}\,yr^{-1}}$, and have an active lifetime\cite{kb17} of $\tau \simeq 10^3$\,yr, magnetars could generate bursts up to a volumetric rate $10^5\,\mathrm{Gpc^{-3}\,yr^{-1}}$, comparable to that of FRBs.}  
\sptvthree{Note that this rate reflects a limiting value of $\alpha\sim-0.15$,
 much \sptvthree{flatter} than those inferred for repeat bursts from FRB 121102\cite{gms+19} ($\alpha=-1.8\pm0.3$) or for repeating FRB 180916\cite{aab+20}
($\alpha=-2.5\pm0.4$).  }

\vk{Moreover,} \spt{\src-like magnetars are unlikely to explain extremely prolific repeating FRBs such as FRB 121102, which has emitted 18 bursts with energies $>10^{37}$\,ergs in 30 minutes\cite{gsp+18}, or be found in the outskirts of elliptical galaxies\cite{bdp+19, pmm+19}. These observations suggest that extragalactic analogues of \vktwo{current} Galactic magnetars could explain some of the FRB population, but 
much more active \vktwo{--- perhaps younger ---} sources, and/or those with non-core-collapse origins,
still need to be invoked to explain all the observations.}


\clearpage

\input{figures}

\input{methods}

\input{extended_data}

\begin{addendum}
\item
\input{acks.tex}
\allacks
    \item[Author Contributions]
    All authors from CHIME/FRB collaboration played either leadership or significant supporting roles in one or more of:  the management, development and construction of the CHIME telescope, the CHIME/FRB instrument and the CHIME/FRB software data pipeline, the commissioning and operations of the CHIME/FRB instrument, the data analysis and preparation of this manuscript.
 	All authors from CHIME collaboration played either leadership or significant supporting roles in the management, development and construction of the CHIME telescope.
    \item[Competing Interests] The authors declare that they have no competing financial interests.
    
    
    \textbf{Correspondence and requests for
        materials} should be addressed to P. Scholz  (email: paul.scholz@dunlap.utoronto.ca).
        
\end{addendum}

\bibliography{NewRefs,frbrefs}

\end{document}

%% file: auth-sgr1935.tex
\author{The CHIME/FRB Collaboration: 
Andersen, B.~C.$^{1,2}$, \allowbreak
Bandura, K.~M.$^{3,4}$, \allowbreak
Bhardwaj, M.$^{1,2}$, \allowbreak
Bij, A.$^{5}$, \allowbreak
Boyce, M.~M.$^{6}$, \allowbreak
Boyle, P.~J.$^{1,2}$, \allowbreak
Brar, C.$^{1,2}$, \allowbreak
Cassanelli, T.$^{7,8}$, \allowbreak
Chawla, P.$^{1,2}$, \allowbreak
Chen, T.$^{9}$, \allowbreak
Cliche, J.~-F.$^{1,2}$, \allowbreak
Cook, A.$^{7,8}$, \allowbreak
Cubranic, D.$^{10}$, \allowbreak
Curtin, A.~P.$^{1,2}$, \allowbreak
Denman, N.~T.$^{11}$, \allowbreak
Dobbs, M.$^{1,2}$, \allowbreak
Dong, F.~Q.$^{10}$, \allowbreak
Fandino, M.$^{10}$, \allowbreak
Fonseca, E.$^{1,2}$, \allowbreak
Gaensler, B.~M.$^{7,8}$, \allowbreak
Giri, U.$^{12,13}$, \allowbreak
Good, D.~C.$^{10}$, \allowbreak
Halpern, M.$^{10}$, \allowbreak
Hill, A.~S.$^{14,15}$, \allowbreak
Hinshaw, G.~F.$^{10}$, \allowbreak
H\"ofer, C.$^{10}$, \allowbreak
Josephy, A.$^{1,2}$, \allowbreak
Kania, J.~W.$^{16}$, \allowbreak
Kaspi, V.~M.$^{1,2}$, \allowbreak
Landecker, T.~L.$^{15}$, \allowbreak
Leung, C.$^{9,17}$, \allowbreak
Li, D.~Z.$^{5,18,19,7}$, \allowbreak
Lin, H.~-H.$^{5,19}$, \allowbreak
Masui, K.~W.$^{9,17}$, \allowbreak
Mckinven, R.$^{7,8}$, \allowbreak
Mena-Parra, J.$^{9}$, \allowbreak
Merryfield, M.$^{1,2}$, \allowbreak
Meyers, B.~W.$^{10}$, \allowbreak
Michilli, D.$^{1,2}$, \allowbreak
Milutinovic, N.$^{10}$, \allowbreak
Mirhosseini, A.$^{10}$, \allowbreak
M\"unchmeyer, M.$^{12}$, \allowbreak
Naidu, A.$^{1,2}$, \allowbreak
Newburgh, L.~B.$^{20}$, \allowbreak
Ng, C.$^{7}$, \allowbreak
Patel, C.$^{7,1}$, \allowbreak
Pen, U.~-L.$^{5,7,21,19,12}$, \allowbreak
Pinsonneault-Marotte, T.$^{10}$, \allowbreak
Pleunis, Z.$^{1,2}$, \allowbreak
Quine, B.~M.$^{22,23}$, \allowbreak
Rafiei-Ravandi, M.$^{12}$, \allowbreak
Rahman, M.$^{7}$, \allowbreak
Ransom, S.~M.$^{24}$, \allowbreak
Renard, A.$^{7}$, \allowbreak
Sanghavi, P.$^{3,4}$, \allowbreak
Scholz, P.$^{7,*}$, \allowbreak
Shaw, J.~R.$^{10}$, \allowbreak
Shin, K.$^{9,17}$, \allowbreak
Siegel, S.~R.$^{1,2}$, \allowbreak
Singh, S.$^{1,2}$, \allowbreak
Smegal, R.~J.$^{10}$, \allowbreak
Smith, K.~M.$^{12}$, \allowbreak
Stairs, I.~H.$^{10}$, \allowbreak
Tan, C.~M.$^{1,2}$, \allowbreak
Tendulkar, S.~P.$^{1,2}$, \allowbreak
Tretyakov, I.$^{7,18}$, \allowbreak
Vanderlinde, K.$^{7,8}$, \allowbreak
Wang, H.$^{9}$, \allowbreak
Wulf, D.$^{1,2}$, \allowbreak
Zwaniga, A.~V.$^{1,2}$ \allowbreak
}
\newcommand{\affils}{
\begin{affiliations}
\item{Department of Physics, McGill University, 3600 rue University, Montr\'eal, QC H3A 2T8, Canada}
\item{McGill Space Institute, McGill University, 3550 rue University, Montr\'eal, QC H3A 2A7, Canada}
\item{CSEE, West Virginia University, Morgantown, WV 26505, USA}
\item{Center for Gravitational Waves and Cosmology, West Virginia University, Morgantown, WV 26505, USA}
\item{Canadian Institute for Theoretical Astrophysics, 60 St.~George Street, Toronto, ON M5S 3H8, Canada}
\item{Department of Physics and Astronomy, University of Manitoba, Allen Building , Winnipeg, MB Canada R3T 2N2}
\item{Dunlap Institute for Astronomy \& Astrophysics, University of Toronto, 50 St.~George Street, Toronto, ON M5S 3H4, Canada}
\item{David A.~Dunlap Department of Astronomy \& Astrophysics, University of Toronto, 50 St.~George Street, Toronto, ON M5S 3H4, Canada}
\item{MIT Kavli Institute for Astrophysics and Space Research, Massachusetts Institute of Technology, 77 Massachusetts Ave, Cambridge, MA 02139, USA}
\item{Department of Physics and Astronomy, University of British Columbia, 6224 Agricultural Road, Vancouver, BC V6T 1Z1, Canada}
\item{Central Development Laboratory, National Radio Astronomy Observatory, 1180 Boxwood Estate Road, Charlottesville VA 22903, USA}
\item{Perimeter Institute for Theoretical Physics, 31 Caroline Street North, Waterloo, ON N2L 2Y5, Canada}
\item{Department of Physics and Astronomy, University of Waterloo, Waterloo, ON N2L 3G1, Canada}
\item{Department of Computer Science, Math, Physics, and Statistics, University of British Columbia, 3187 University Way, Kelowna, BC V1V 1V7, Canada}
\item{National Research Council Canada, Herzberg Research Centre for Astronomy and Astrophysics, Dominion Radio Astrophysical Observatory, PO Box 248, Penticton BC Canada V2A 6J9}
\item{Department of Physics and Astronomy, West Virginia Univeristy, Morgantown WV 26506, USA}
\item{Department of Physics, Massachusetts Institute of Technology, 77 Massachusetts Ave, Cambridge, MA 02139, USA}
\item{Department of Physics, University of Toronto, 60 St.~George Street, Toronto, ON M5S 1A7, Canada}
\item{Max-Planck-Institut für Radioastronomie, Auf dem Hügel 69, Bonn, 53121, Germany}
\item{Department of Physics, Yale University, New Haven, CT 06520, USA, USA}
\item{Canadian Institute for Advanced Research, CIFAR Program in Gravitation and Cosmology, 661 University Ave, Toronto, ON M5G 1Z8, Canada}
\item{Thoth Technology Inc., Algonquin Radio Observatory, Achray Rd, RR6, Pembroke, ON K8A 6W7, Canada}
\item{Department of Physics and Astronomy, York University, 4700 Keele St, Toronto Ontario, M5S 2M7, Canada}
\item{National Radio Astronomy Observatory, 520 Edgemont Road, Charlottesville, VA 22903, USA}
\end{affiliations}
}

%% file: figures.tex
\begin{figure}[htbp]
\begin{center}
\includegraphics[width=\textwidth]{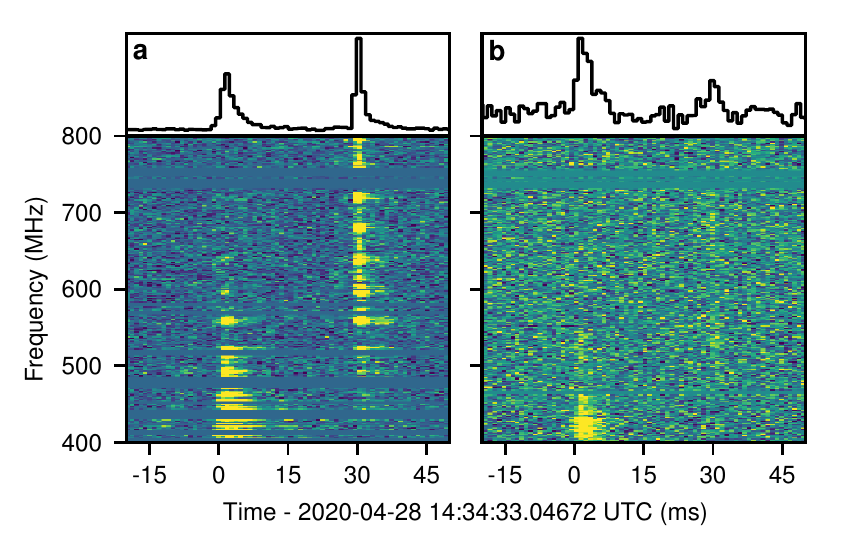}
%
\caption{
\textbf{Burst waterfalls.} {Total intensity normalized dynamic spectra and band-averaged time-series (\postsub{400.1953125-MHz arrival time} referenced to the geocentre) of the detections by (a) CHIME/FRB and (b) ARO, relative to the geocentric best-fit arrival time of the first sub-burst based on CHIME/FRB data. \psvtwo{For CHIME/FRB, the highest S/N beam detection is shown.} Dynamic spectra are displayed at 0.98304-ms and 1.5625-MHz resolution, with intensity values capped at the 1st and 99th percentiles. Frequency channels masked due to radio frequency interference are replaced with the median value of the off-burst region. The CHIME/FRB bursts show \psnote{a ``comb-like'' spectral structure due to} their detection in a beam sidelobe as well as dispersed spectral leakage that has an instrumental origin (see Methods).}
}
\label{fig:waterfall}
\end{center}
\end{figure}

\clearpage
\begin{figure}[htbp]
\begin{center}
\includegraphics[width=\textwidth]{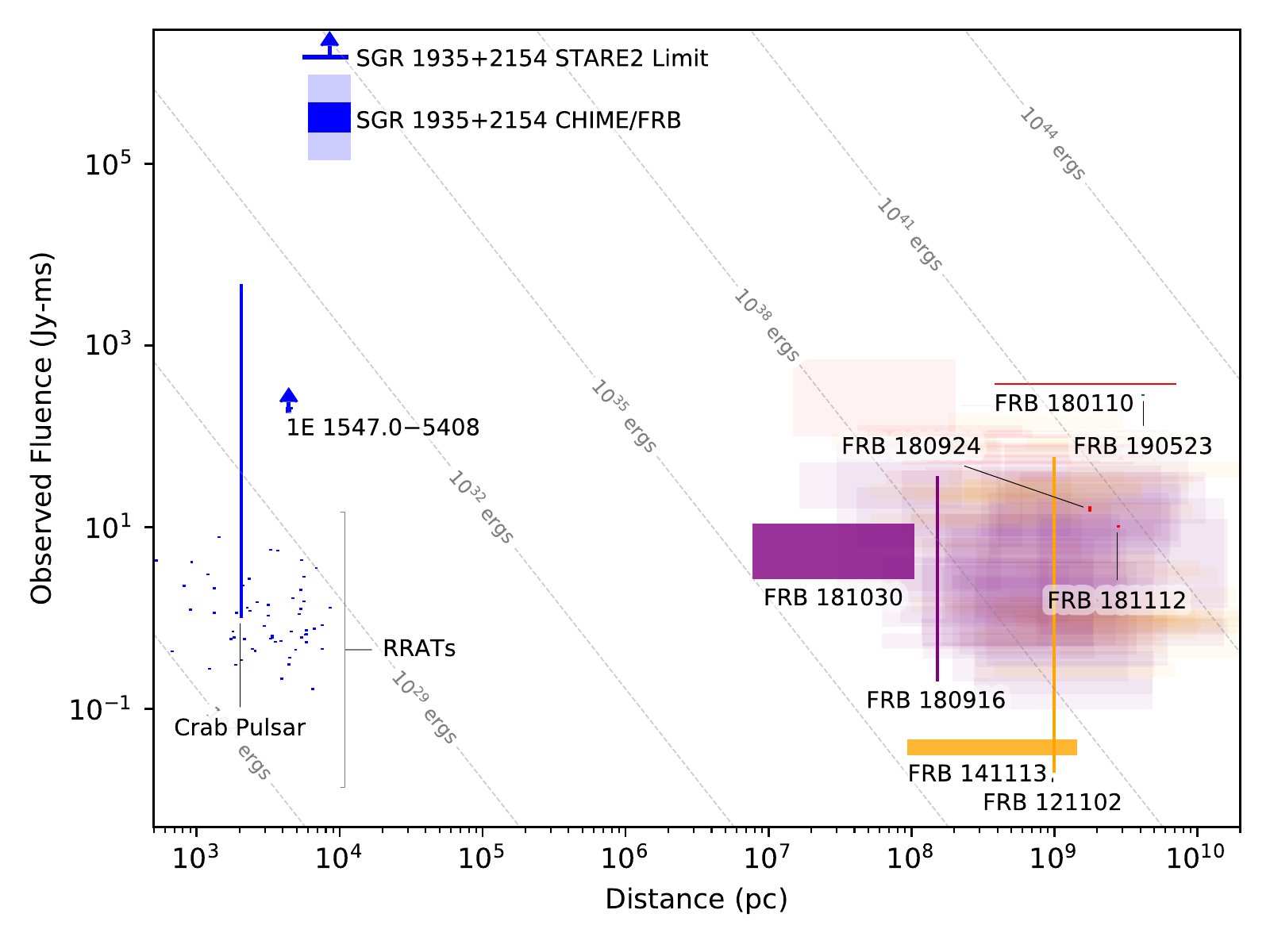}
%
\caption{
\textbf{Comparison of short radio burst energetics.} The observed burst fluences at radio frequencies from 300\,MHz to 1.5\,GHz for Galactic neutron stars and extragalactic FRBs are plotted with their estimated distances. The fluence ranges include the uncertainties in fluence measurements as well as ranges of individual bursts for repeating FRBs and pulsars. FRBs colours indicate their detection telescope: CHIME/FRB (purple), ASKAP (red), DSA-10 (green, \sptvthree{FRB\,190523}), Arecibo and Parkes (orange). Galactic sources are plotted in blue. For \src\ the blue rectangle indicates the nominal range of \vktwo{400--800-MHz} fluences measured for the two bursts while the light blue region incorporates the possible systematic uncertainty in the CHIME/FRB fluence as described in the text. \sptvthree{The STARE2 lower limit on the fluence at 1.4\,GHz is also shown.} Gray diagonal lines indicate loci of equal isotropic burst energy with an \sptvthree{assumed fiducial} bandwidth of 500 MHz. FRB distances are estimated from their extragalactic dispersion measure contribution including the simulated variance \cite{plm+19}. Pulsar distances are estimated based on the NE2001 Galactic electron distribution model \cite{cl02}. Objects with accurately measured distances (parallax or host galaxy redshift) are indicated with vertical lines. 
}
\label{fig:burstenergies}
\end{center}
\end{figure}

\clearpage

%% file: methods.tex
\begin{methods}

\subsection{Localization}

To determine the sky location of the event, we fit a model for the CHIME/FRB synthesized beam and underlying Gaussian burst spectrum to the measured spectra \psnote{of the second, brighter burst component. The free parameters in the model were the sky position and the width, mean, and amplitude of the Gaussian spectral model. Sixty-eight burst spectra were fitted: 54 consisting of the main cluster of beams that detected the event (out of a total of 93 detection beams) plus 14 adjacent non-detection beams.}
The fitting was performed using a Markov Chain Monte Carlo method\cite{fhlg13} with a flat prior on the position of the event in a $108^\circ\mathrm{(E-W)}\times10^\circ\mathrm{(N-S)}$ region centred on the CHIME meridian at the zenith angle of the beam with the brightest detection.
In this large area of the sky searched, the localization optimization settled on a position of RA = $293.9^\circ$, Dec = 22.1$^\circ$, which is $0.3^\circ$ from the position of \src. As this localization method is still in development, and is largely untested for events this far off the CHIME meridian, we believe there are significant as-yet unquantified systematic uncertainties of order $1^\circ$ and so do not \revMD{report the}
statistical uncertainties on the fit, \revMD{which is subdominant}. 

\subsection{Estimate of Burst Fluence}

The central challenge of obtaining a flux calibration for the burst is characterizing the primary beam response far outside its main lobe, \revMD{where the telescope was optimized for detection and calibration.}
The CHIME primary beam main lobe covers a narrow strip along the meridian
\revMD{centered at hour angle HA=0 and}
2--3 degrees wide depending on frequency, 
\revMD{and extends about $\pm$60 degrees in elevation.} When the burst was detected, \src\ was at HA = 22$^{\circ}$.   We measure the primary beam response at this location using both interferometric 
measurements \revMD{of the sun} and holographic measurements of the Crab nebula. Although both measurements are fundamentally interferometric in nature, they are subject to different systematics --- not least of which is the use of different celestial sources (i.e., the Sun versus the Crab). As will be explained in more detail below, the two measurements are made using different antenna-baseline configurations, and use fairly independent gain normalization schemes. 

The proximity in declination of the Crab Nebula to \src\ 
\revMD{ provides a fortuitous means of calibrating the primary beam at that declination as a function of hour angle.}
The Crab Nebula is located roughly 0.12$^{\circ}$ \vk{North} in declination
from the \src, which induces a systematic error at the few times 10\% level \kwm{(estimated from the solar data described below)}.
Because \src\ is located 
\revMD{HA=}\vk{22$^{\circ}$}
from meridian, far into the sidelobes of the primary beam, interferometric data from the CHIME telescope alone contains too much confusion noise from other radio sources to estimate the beam using the Crab Nebula. To eliminate confusion with other sources, and to boost the signal in this low-sensitivity part of the beam, we employ a version of the standard radio holographic technique. We track the Crab Nebula using the co-located DRAO 26-m Galt telescope, outfitted with a 400--800\,MHz  receiver, and perform interferometry with the stationary CHIME antennas \cite{naa+14, bna+16}. The resulting data provide a measurement of the complex response of each antenna and polarization in the CHIME array along the source transit path in hour-angle, which contains only correlated power between the primary CHIME telescope beam and the primary Galt telescope beam. In traditional radio holography, such a measurement would then be transformed to make a map of the aperture of the telescope, usually to assess telescope surface figure. 

The quantity we wish to measure is the complex beam response to an unpolarized source, averaged over the array of feeds and summed in intensity over polarization channels. Following the formalism in \cite{msn+19}, we write this as\footnote{Here we suppress the spectral frequency axis and the beam measurement should be understood to be performed independently in each spectral channel.}

\begin{equation}
    B(\theta) \equiv \frac{1}{n_{\rm ant}}\sum_p\left| \sum_i \vec{A}_{ip}(\theta)\right|^2,
    \label{e:pribeam}
\end{equation}
where $\vec{A}_{ip}(\theta)$ is the direction-dependent response to the vector electric field of feed $i$ and polarization channel $p$,
and $n_{\rm ant}$ is the number of feeds. In this case, we have made a measurement at a single declination and thus $\theta$ is simply the hour-angle during the transit of the Crab Nebula.
The measured holographic visibility can be used to estimate $\vec{A}_{ip}(\theta)$ for CHIME, and is defined as:
\begin{equation}\label{eq:holo_vis}
    V_{ip, p'}^{\mathrm{meas}}(\theta) = g^{\rm C}_{ip} g^{\rm Galt}_{p'}\vec{\epsilon}_{p'} \cdot \vec{A}_{ip}(\theta) S_\nu,
\end{equation}
where $g^{\rm Galt}_{p'}$ is the overall gain of the polarization channel $p'$ of the Galt Telescope (including optical, analogue, and digital contributions), $\vec{\epsilon}_{p'}$ is the unit vector denoting the direction of the channel's polarization response (we assume these to be orthogonal for the two Galt polarization channels), $g^{\rm C}_{ip}$ is a per-feed gain factor for the CHIME telescope (analogue, and digital contributions), set to a real number by phase-referencing at the Crab Nebula transit, and $S_\nu$ is the frequency-dependent source flux density of the Crab Nebula in Janskys. 

To use the holographic visibilities as an estimate of the CHIME beam, we must account for a few scalings: 

First, the holographic visibility for a given CHIME polarization channel, $p$, is normalized by the total power measured on meridian. This term is the sum of the powers in the co-polarization and cross-polarization between CHIME and the Galt telescope: $\sqrt{ \sum_{p'} |V_{ip,p'}(0)|^2 }$.

Second, the gains between the two polarization channels on the Galt telescope are not equal. We can calibrate the relative gains between the two polarizations using the Galt auto-correlations: 
\begin{equation}
    V_{p, p}^{\mathrm{Galt}} \propto |g^{\rm Galt}_{p}|^2 S_\nu,
\end{equation}
We have neglected the contribution from the receiver temperature due to the brightness of the Crab nebula and the relatively high gain of the Galt telescope.
Dividing by the square-root of the auto correlations removes the dependence on the relative gains of the two polarizations.
This modifies the normalization factor derived from the total power above, as: $$ \sqrt{V_{p', p'} \sum_{p''} \left[ |V_{ip,p''}(0)|^2/ V_{p'',p''} \right]}  .$$

Third, the gain for each CHIME feed on meridian is computed and stored for bright transits, including the Crab Nebula. We can use this to scale the holographic data, per feed and polarization, to the CHIME response values at meridian: $\vec{A}_{ip}(0)$, which also accounts for the $g_{ip}^{C}$ term in Equation~\ref{eq:holo_vis}. 

Provided the direction of $\vec{A}_{ip}(0)$ does not depend on $i$ (which is empirically true at the 1\% level), the desired beam response can then be written as:
\begin{equation}
    B(\theta) = \frac{1}{n_{\rm ant}}\sum_{p,p'}\left|\sum_i \frac{V_{ip, p'}(\theta)\vec{A}_{ip}(0)}{\sqrt{V_{p', p'} \sum_{p''} \left[ |V_{ip,p''}(0)|^2/ V_{p'', p''}\right]}}\right|^2.
\end{equation}
Finally, we compensate for hour-angle dependent reduction in the CHIME-Galt correlation amplitude due to the $\sim 1500$\,ns delay of the arrival of the Galt telescope signals at the correletor relative to the signals from the CHIME antennas.
We evaluate this response at the position of closest approach of the Crab to the burst location in the CHIME beam using data collected January 24, 2019. In rough terms, we find that the primary beam at this location has 0.5\% of the meridian response in the lower half of the band\dww{, falling to}
0.1\% of the meridian response in the top half of the band. \dww{Beyond this general trend, the primary beam response across the band is dominated by spectral structure on $\sim$30~MHz scales due to multiple reflections within the telescope.}

\src{} is in the declination range that is seasonally covered by the Sun. An empirical beam model covering the full solar declination range was built using CHIME interferometric data from 2018 and 2019. \revRS{To construct the model, we beamform to the location of the Sun using the purely North-South} (same-cylinder) baselines.
This average thus includes feed-to-feed variations in both amplitude and decoherence due to variations in phase.
\revMD{We expect these variations to be statistically representative, although not with the identical feed weights as the FRB beamformer.}
The solar data is also analytically corrected for digital clipping which occurs due to the limited bit depth of the correlator.
Separate beam models are constructed for the East-West and North-South antenna polarizations, which are then summed. The flux of the Sun is \revMD{cross-}calibrated \revMD{with the Crab Nebula}, 
when \dww{they share a common declination.}
While the flux of the Sun can be quite variable, the proximity in declination of \src{} to the Crab, and that the data were collected near solar minimum, \dww{reduce the likelihood that the solar flux is varying significantly.}
\dww{We do not attempt to correct for the spatial extent of the Sun, which is $\sim 0.5^\circ$ in diameter and therefore resolved by the longest baselines.  Averaged across the band, we roughly estimate that this effect accounts for a $20\%$ reduction in observed solar flux.}
However, since the baselines are in the North-South direction, the fraction of resolved flux should be nearly constant between the beam location of the burst and calibration with the Crab. The resulting primary beam has qualitatively similar structure to that measured from holography, but an amplitude roughly a factor of two lower at the observed location of the burst (implying a higher intrinsic flux of the \src{} burst). 
\revMD{This difference between solar and holography calibration methods is larger than expected and is taken as a systematic contribution to the flux measurement.}

We further verify these beam response measurements using beamformed data for the Crab Nebula acquired through the FRB backend. This measurement is more direct than combining interferometric measurements of the primary beam with a model for the formed beams. However, the measurement is noisy since it suffers from source confusion as other sources contribute significant flux when the Crab is attenuated by the sub-percent level beam response. Baseline subtraction is also challenging since we cannot slew the sidelobe to an off-source location. Nonetheless, the Crab is clearly detected even at hour angle 22$^{\circ}$ and validates our beam measurements at the few tens of percent level, in better agreement of that obtained with holography than solar data.

\revRS{We proceed with flux and fluence determination using measurements of the primary beam from both holography and solar data. Our quoted flux densities and fluences are the averages of those obtained with the two primary beam models. As the beam responses from the three measurements agree only at the factor-of-two level, we
conservatively ascribe a factor-of-two systematic error to both the flux density and fluence estimates to account for this disagreement. Further analysis and observations will be required to 
\revMD{improve the far sidelobe calibration accuracy.}}

\bca{For each beam in which \src{} was detected, we calculate beam-attenuated fluence spectra for both subbursts by integrating over the extent of each burst in the CHIME/FRB dynamic spectrum. We derived peak flux spectra by multiplying the fluence spectra by the fluence-to-peak-flux ratio for each frequency channel calculated from the best-fit model of the intrinsic burst spectrum (described in the next subsection). 
Absolute flux scaling is derived from the real-time array calibration that is determined daily from bright-source transits and applied prior to beamforming (accounting for scaling factors inherent to the beamforming algorithm).
In addition, we multiply our final flux density values by a factor of 1.09 to account for the flux aliased outside the burst extent due to spectral leakage during the frequency channelization process.}

We fit the measured fluence and peak flux spectral data to a model including the primary beam, formed beam, and burst spectrum. We model the composite beam as the product of the measured primary beam and formed beams. Since our primary beam measurements are feed-averages of the complex primary beam (see Equation~\ref{e:pribeam}), our model accounts for reductions in the sensitivity of the synthesized beam due to beam phase variations, which, at this hour angle, is a $\sim20\%$ effect at the bottom of the band and an order unity effect at the top of the band. Beam phase variations also cause an \emph{increase} in the formed beam sensitivity at frequencies where they would have a null response in the absence of phase variations, and this effect is not accounted for. As such, in the fits we only include spectra for the 4 beams with the largest response, where the fits are dominated by frequencies of high formed-beam sensitivity.
We model the intrinsic spectrum as a power law with a spectral running (see below). The model also includes two free parameters to shift the positions of the grid of formed beams relative to the source. This is necessary because \psnote{of a known systematic drift between the formed beam model and the true position of the formed beams as a function of distance from the CHIME meridian which is currently under investigation.} The fit yields a position offset of 0.1$^\circ$, and is nearly identical between fits to the first and second sub-bursts. We use least-squares with equal weighting to fit the spectra from the four beams to the data. For our reported fluences and fluxes, we average the best-fit spectral model over 400 to 800\,MHz.

\subsection{Burst morphology and spectra}

We analyzed the CHIME/FRB dynamic spectrum using a multi-component modeling scheme and least-squares algorithm described in previous  works\cite[e.g.]{abb+19c}. The model spectrum consists of two distinct components, assumed to be intrinsically Gaussian in \psnote{temporal} shape. Both of the model components have identical dispersion and scattering properties applied (i.e., dedispersed to the same DM, and scatter-broadened by the same one-sided exponential pulse broadening function). Each component has an independently determined arrival time, amplitude, spectral shape, and temporal width that is corrected for broadening from single-tail scattering and dispersion smearing. The spectral energy distribution as a function of frequency, $I(f)$, is assumed to follow a weighted power-law form,

\[
    I(f) \propto (f/f_0)^{\gamma + r\ln(f/f_0)},
\]

\noindent with the spectral index ($\gamma$) and ``running" of the spectral index ($r$) as free parameters when modeling the dynamic spectrum, while the reference frequency ($f_0$) is held fixed to an arbitrary value. Moreover, the dispersion delay is $\Delta_{\rm DM} = k\textrm{DM}f^{-2}$, with $k = 4149.377 \textrm{ s pc}^{-1}\textrm{ cm}^3\textrm{ MHz}^2$, while the scattering timescale is assumed to be proportional to $f^{-4}$.

Unlike previous analyses, the two-component model used for parameter estimation was weighted by the frequency-dependent beam response in the direction of \src{} described in the preceding subsection. Such weighting accounts for spectral effects due to sidelobe detection, and thus allows for direct modeling of the ``intrinsic" (i.e., beam-corrected) dynamic spectrum. Use of the beam-response model during the least-squares estimation of burst parameters is statistically preferred over models that do not explicitly account for the instrumental fringe pattern. No attempt was made to directly account for the faint, ``ghost" leakage artifacts in the observed spectrum when fitting for the best-fit model, and will be the subject of future work.

The intrinsic and beam-attenuated models, as well as the observed dynamic spectrum and best-fit residuals (i.e., differences between the model and data) are shown in Extended Data Figure \ref{fig:fitburst}. Remaining, non-zero structure in the residuals likely reflects departures from the simplistic assumptions made for describing the spectral energy distribution. The estimated parameters that describe the best-fit model of the spectrum are presented in Table \ref{tab:properties}.

\subsection{Algonquin Park 10-m Radio Telescope Observations}

The Algonquin Radio Observatory (ARO) hosts a stationary 10-m single-dish telescope in Algonquin Provincial Park, Ontario, Canada. The telescope uses a cloverleaf feed identical to those on CHIME\cite{baa+14}, utilizing an ICEboard as a digital backend\cite{bbc+16}, with a frequency range of 400--800 MHz using the same native 1024-bin channelization as CHIME. 
The system is continuously recording baseband data to a rolling 24.5-hr disk buffer
and was pointed at (RA, Dec) = ($318^{\circ}$, $22^{\circ}$) at the time of the radio burst from \src. The baseband data was manually saved after the CHIME event, which
in the future should occur automatically.

Based on the pointing and localization from the CHIME
detection, the event was in a far sidelobe of the ARO beam. The
burst was identified in the data with the CHIME-measured timing.
Each linear polarization was coherently dedispersed with DM = 332.80 pc~cm$^{-3}$, and the Stokes Q, U, and V parameters were formed subtracting the off-pulse component. The off-pulse per-channel RMS values are used as weights for the polarization analysis \cite{fab+20}. 

The first burst (B1) is detected in both polarizations. We use RM synthesis to measure its RM from the Stokes Q and U parameters, after correcting instrumental leakages. The resulting Faraday spectrum is shown in Extended Data \ref{fig:fspec}a, contributing to final RM = 116 rad~m$^{-2}$, see below. 

The second burst (B2) is only visible in one linear polarization in the ARO data (which we label $Y$) due to frequency variations in the polarized sidelobe response. To account for this, we infer the RM through the flux in the polarized flux \yy, which is related to the Q and I stokes parameters by \yy$=(I+Q)/2$. 
We perform RM Synthesis \cite{burn66,bd05} on \yy. 
We identify a peak in the amplitude of the Faraday spectra at $\mr{RM}\sim116$ rad~m$^{-2}$ (Extended Data \ref{fig:fspec}b). The RM measured in this way will be free from the influence of U-V leakage and have different systematics compared to the measurement using Q and U.

We verify this method using two bright giant pulses from the Crab Pulsar detected with the same instrument. We are able to recover identical values of RM and PA difference from the single linear polarization as those obtained from combining Stokes Q and U, although with half the significance.  Moreover, we compare the Faraday spectra from \yy\ only with the traditional one from combining Q and U for B1. The offset between the measured RM and polarization angle (PA) from those approaches are 0.7 rad~m$^{-2}$ and 1.6$^{\circ}$, which are well within the uncertainties.  The crab data are also used to determine the sign of RM\cite{sbg+19}.

We cross correlate the Q-U Faraday spectra $F_\mr{B1}$ from the first burst with the \yy\ Faraday spectra $F_\mr{B2}$ from the second burst $F_\mr{cross}=\sqrt{F_\mr{B1} F^*_\mr{B2}}$. Assuming no RM changes between the two bursts, separated by only $29$\,ms, the phase of the cross spectrum corresponds to the PA difference between the two bursts. A maximum cross-correlation signal appears at RM$=116\pm2$ ($\sigma_\mr{measurement}$) $\pm5$ ($\sigma_\mr{systematic}$) rad~m$^{-2}$, with a PA difference of $5\pm10$ deg between the bursts at the peak. As shown in Extended Data Figure~\ref{fig:fspec}c, near the peak, the real part of the cross spectrum dominates the amplitude, indicating the PA difference is small, consistent with zero. The systematic uncertainty $\sigma_\mathrm{systematic}$ is estimated from the change of inferred RM due to different Stokes U-V leakage models, that arise from a combination of side lobe phases and differential cable delays, as well as uncertainties of the underlying spectral shapes.

Although CHIME has only recorded intensity data, the oscillations of Stokes Q due to Faraday rotation can leak into the summed intensity through the different responses of the two linear polarization feeds in the far sidelobe. Consequently, we can estimate the RM and PA difference using a similar procedure to the baseband analysis. We see a signal near RM$\sim116$ for both bursts. The cross spectrum is shown in Extended Data Figure~\ref{fig:fspec}d, with a peak at $115.3$) rad~m$^{-2}$. The measured PA difference at the peak of the cross spectrum is $4\pm10$ deg, consistent with the result from the ARO 10m.

Extended Data Figure~\ref{fig:polarization} shows the model of fitted PA and RM against data for B1, where the characteristic signal of Faraday rotation is identified. The measured RM is similar to that reported by FAST (RM = 112.3 rad~m$^{-2}$) for a different burst from the same source\cite{zjm+20}. Taken together, these factors indicate that the ARO measured RM is robust in spite of the unknown systematics. Robust linear and fractional polarization measurements await calibration with a beam model that has been validated with bright sidelobe events of known sources.

\subsection{CHIME/FRB Non-Detections of November 2019 High-Energy Bursts from \src}

\vk{\src\ is in an extended active phase that commenced in Fall 2019, and which has included many dozens of X-ray and gamma-ray  bursts.  Four high-energy bursts
occurred in November 2019, while the source was closer to CHIME's meridian than for the 28 April 2020 burst.  Specifically, bursts on
4 November 2019 at UTC 01:20:24\cite{gcn26163} seen by the {\it Fermi} Gamma-ray Burst Monitor, and UTC 01:54:37\cite{gcn26169} seen by the {\it Swift} Burst Alert Telescope (BAT), as well as on 5 November 2019 at UTCs 00:08:58\cite{gcn26169} and 01:36:25\cite{gcn26171}, both seen by {\it Swift}/BAT, were
at CHIME hour angles 9.5$^{\circ}$, 18.1$^{\circ}$, $-7$.4$^{\circ}$, and 14.5$^{\circ}$, respectively, when they occurred.  All four occurred during nominal CHIME/FRB operations.} 
No radio burst was detected by our automated pipeline at any of the four high-energy burst epochs.
\psvtwo{(We also ran a custom implementation of the clustering algorithm DBSCAN\cite{eksx96} tailored to the CHIME/FRB pipeline to identify any other events from this source at any epoch, including in
 the sidelobes but found no detections.)}
Using our current best models to account for CHIME’s primary beam response, the formed beam response, and system sensitivity, (corroborated using techniques analogous to those described above), and scaling the measured fluences for three detected FRBs at nearly the same declination as \src\
using their detection signal-to-noise ratios and the CHIME/FRB detection threshold of 10,
we determine a conservative upper limit on radio burst flux density in the 400--800 MHz range of $<1.7$~kJy,
$<4.6$~kJy, $<1.2$~kJy, and $<3.0$~kJy
for the four events, respectively, \psvtwo{within an approximately 3-$\sigma$ confidence interval, though the reader is cautioned that the challenges of working in the far sidelobe make a precise confidence interval difficult to calculate.}
\vk{Thus, radio bursts at the epochs of these high-energy events from \src\  had 400--800-MHz radio flux densities factors of at least 30--120 times lower than the 28 April 2020 event. }

For the two bursts on 5 November 2019, we determine the X-ray flux from the processed online public BAT data products using Version 12.10.1f of HEASARC’s XSPEC program\footnote{https://heasarc.gsfc.nasa.gov/xanadu/xspec/}. 
While the high-energy fluxes for the bursts on 4 November 2019 will be considered in future work, 
we note that the {\it Swift}/BAT burst at 01:54:37 on 4 November is much weaker than the two {\it Swift}/BAT bursts on 5 November.
For the burst at UTC 00:08:58, we find the data are well modelled using two blackbody components. Fixing the equivalent neutral hydrogen absorption column\cite{ier+16} $N_H$ to $1.6 \times 10^{22}$~cm$^{-2}$, we find a reduced $\chi^2$ of 1.0004 for this model. The 
20--200-keV flux is then $6.7^{+0.3}_{-0.6}
\times 10^{-7}$~erg~cm$^2$~s$^{-1}$.
For the November 5 burst at UTC 01:36:25, we follow a similar procedure and use a double blackbody model to find reduced $\chi^2 = 0.75$. The 
20--200-keV flux is then $8.3^{+0.2}_{-0.2}
\times 10^{-7}$~erg~cm$^2$~s$^{-1}$. 
\vk{Thus, both 5 November 2019 bursts had X-ray fluxes 14 and 11 times, respectively, \revMD{smaller} than the reported Konus-{\em Wind} 20-200~keV high-energy flux\cite{rga+20} on 28 April 2020 ($9.1 \pm 2.6
\times 10^{-6}$~erg~cm$^2$~s$^{-1}$),
in contrast to being respectively \revMD{more than} 120 and 50 times smaller in the CHIME band.  This shows the radio to X-ray flux ratios for these 2019 events were at least 9 and 4 times smaller, respectively, than for
the 28 April 2020 event.}

\subsection{Comparison between volumetric rates of magnetar flares and FRBs}

We estimate the total number of active magnetars in the CHIME-visible volume by scaling the total star-formation rate in nearby galaxies to that of the Milky Way. From the complete sample of galaxies\cite{kk13} within 11\,Mpc, \vk{the distance to which the April 28 \src\ burst, at its best-estimated 400--800-MHz flux density, would have been seen by CHIME/FRB}, we selected 15 galaxies in the CHIME field of view (declination $> -10^\circ$) that had high star-formation rates and large CHIME exposure (Extended Data Table~\ref{tab:nearby_galaxies}). The total star-formation rate in these galaxies is $\approx 36\,\mathrm{M_\odot\,yr^{-1}}$ based on H$\alpha$ intensity estimates and total FUV luminosity (where H$\alpha$ estimates were not available) and far infrared luminosity\cite{jcb+19} for M82 ($\gtrsim10\mathrm{M_\odot\,yr^{-1}}$). Given the star-formation rate in the Milky Way ($\approx1\,\mathrm{M_\odot\,yr^{-1}}$) and the population of $\approx$ 30 active magnetars\cite{ok14}, we estimate about $10^3$ active magnetars exist in this volume.

During CHIME/FRB's operations since September 2018, we have not observed an FRB above a signal to noise ratio of 9 from any of these galaxies. Given this non-detection, we can set an upper limit on the rate of \src-like bursts from any magnetar \vk{in this range}. We estimated the exposure time of CHIME/FRB to these galaxies and calculated the sum of the exposure times of all the galaxies weighted by their star-formation rate (and hence the number of active magnetars). We estimate that the total exposure of CHIME/FRB was $6\times10^{4}\,$magnetar-hours.

The 95\% confidence upper limit\cite{geh86} on the rate of \src-like bursts with energy $E>10^{34}\,\mathrm{erg}$ per magnetar is $R_B(E>10^{34}\,\mathrm{erg}) <5\times 10^{-5}\,\mathrm{hr^{-1}} \approx 0.4\,\mathrm{yr^{-1}}$.

The lower limit on the rate of magnetar bursts in the galaxy comes from the single detection of a burst from \src. Ten of the thirty Galactic magnetars pass over the CHIME primary beam. A bright burst like the one from \src\ would be detectable by CHIME if the source was above the horizon. With an on-sky time of one and a half years since September 2018, we can estimate a 95\% lower limit\cite{geh86} on the rate to be $R_B(E>10^{34}\,\mathrm{erg}) > 7\times10^{-3}\,\mathrm{yr^{-1}}$.

The ASKAP telescope searched for bursts from the Virgo Cluster at 16.5 Mpc with a 300-hr long observation\cite{alf+19} with a 10-$\sigma$ sensitivity of $\approx30\,$Jy\,ms. At a distance of the Virgo cluster and with the observing bandwidth of 336\,MHz, this fluence threshold corresponds a burst energy $\sim10^{36}$\,ergs. The star-formation rate in the Virgo cluster is estimated to be $776\,\mathrm{M_\odot\,yr^{-1}}$, corresponding to about $2.3\times10^4$ \src-like magnetars. Thus, the upper limit on the rate of magnetar bursts is $R_B(E>10^{36}\,\mathrm{ergs}) < 4\times10^{-7}\,\mathrm{hr^{-1}}\, \approx4\times10^{-3}\,\mathrm{yr^{-1}}$.

\end{methods}

%% file: extended_data.tex

\begin{extended-data}

\begin{table} 
\caption{\textbf{Nearby Starforming Galaxies}}
\begin{center}
\begin{tabular}{lccccc}
\hline\hline
Name & RA & Dec & Distance & SFR & Exposure$^1$ \\
     & (deg) & (deg) & (Mpc) & ($\mathrm{M_\odot\,yr^{-1}}$) &  (hrs) \\
\hline
NGC 4559     &	188.99 &	27.96 &	7.4  &	1.1	    &35\\
NGC 4490     &	187.65 &	41.64 &	6.9  &	0.79	&51\\
NGC 1569     &	67.70  &	64.85 &	3.2  &	0.78	&58\\
UGCA 127     &	95.23  &	-8.50 &	10.0 &	1.3 	&40\\
NGC 4258     &	184.74 &	47.30 &	7.7  &	2.9 	&21\\
NGC 3556     &	167.88 &	55.67 &	9.6  &	0.91	&68\\
NGC 5194     &	202.47 &	47.23 &	7.6  &	3.0 	&21\\
M81          &	148.89 &	69.07 &	3.7  &	0.98	&66\\
NGC 2903     &	143.04 &	21.50 &	9.2  &	1.9 	&38\\
NGC 3521     &	166.45 &	-0.04 &	8.5  &	3.2 	&23\\
NGC 3627     &	170.06 &	12.99 &	8.5  &	2.3 	&32\\
NGC 5055     &	198.96 &	42.03 &	9.0  &	1.6 	&50\\
M101         &	210.80 &	54.35 &	6.95 &	3.3 	&40\\
NGC 6946     &	308.71 &	60.15 &	7.7  &	2.5 	&74\\
M82          &	148.97 &	69.68 &	3.6  &	10  	&94\\
\hline
\end{tabular}
\end{center}
\label{tab:nearby_galaxies}
$^1$: Exposure in the full width at half maximum of CHIME beams.
\end{table}

\begin{figure}[htbp]
\begin{center}
\includegraphics{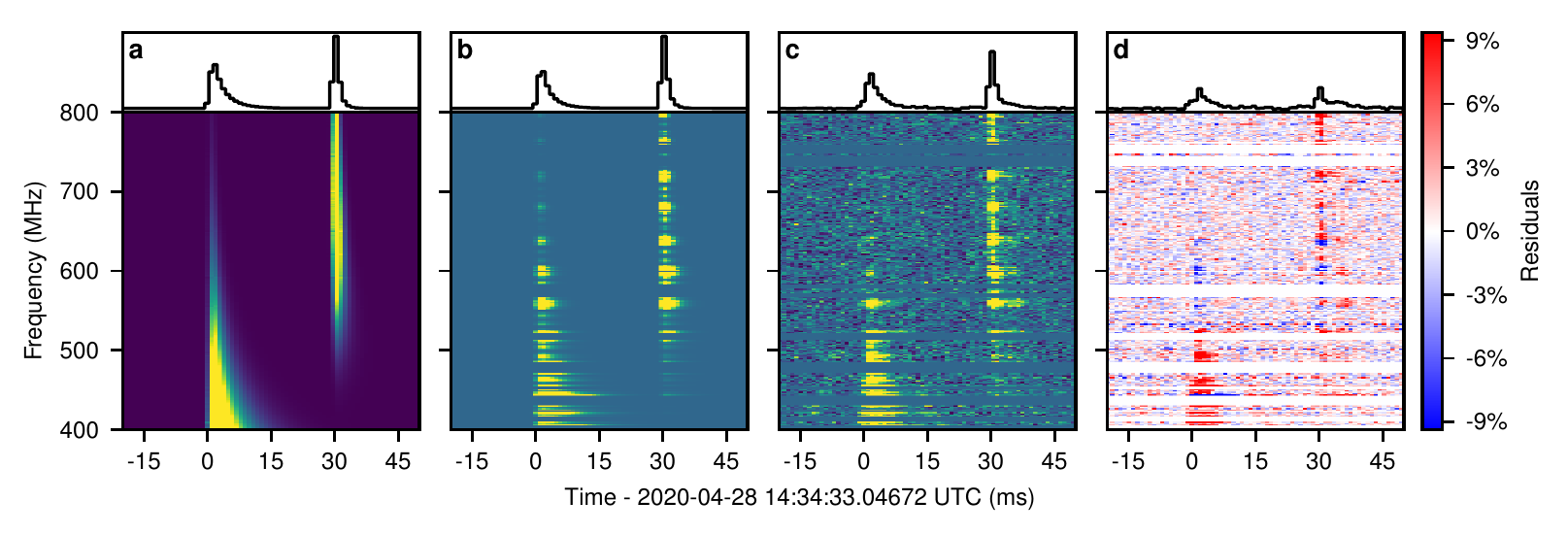}
%
\caption{
\textbf{Burst fitting.} Dynamic spectra and band-averaged time-series (referenced to the geocentre) of (a) fitted burst models, (b) beam-attenuated burst models, (c) burst data as in Fig.~\ref{fig:waterfall}) and (d) fit residuals. Dynamic spectra are displayed at 0.98304-ms and 1.5625-MHz resolution, with intensity values capped at the 1st and 99th percentiles, except in (d) where values are capped at $\pm3\sigma$ around 0. The time-series of (b)--(d) have the same scaling. The beam-attenuation of the maxima in the model dynamic spectra is about $1700\times$.
}
\label{fig:fitburst}
\end{center}
\end{figure}

\clearpage
\begin{figure}[htbp]
\begin{center}
\includegraphics[width=\textwidth]{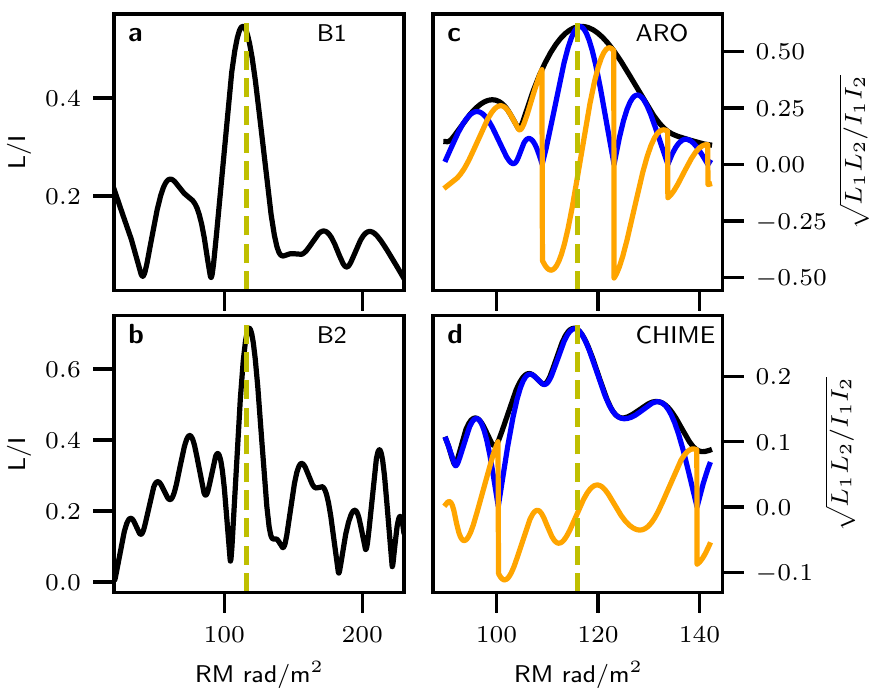}
%
\caption{
\textbf{Polarized intensity Faraday spectra for the two bursts.}
\textbf{a}: The Faraday spectrum $F_\mr{B1}$ for the first sub-burst from Stokes Q and U after correcting a leakage between Stokes U and V. 
\textbf{b}: Faraday spectrum $F^*_\mr{B2}$ for the second sub-burst from a single polarized flux of ARO 10m dish;
\textbf{c}: The cross spectrum $F_\mr{cross}=\sqrt{F_\mr{B1} F^*_\mr{B2}}$ from ARO 10-m dish zoomed in near the peak; 
\textbf{d}: The cross spectrum from CHIME intensity data. The oscillations of Stokes Q from Faraday rotation has leaked to the summed intensity due to different response of the two linear receiver in the far sidelobe. 
The dark lines show the amplitude of the spectra, while the blue and orange lines are real and imaginary part of the spectra, respectively. 
The phase of the cross spectrum corresponds to the PA difference between the two bursts. When the real part approaches the amplitude, the two bursts have the same PA.
The yellow vertical line is drawn at RM$=116$\,rad\,m$^{-2}$. 
}
\label{fig:fspec}
\end{center}
\end{figure}
\clearpage

\clearpage
\begin{figure} [t!]
\begin{center}
\includegraphics[scale=0.3]{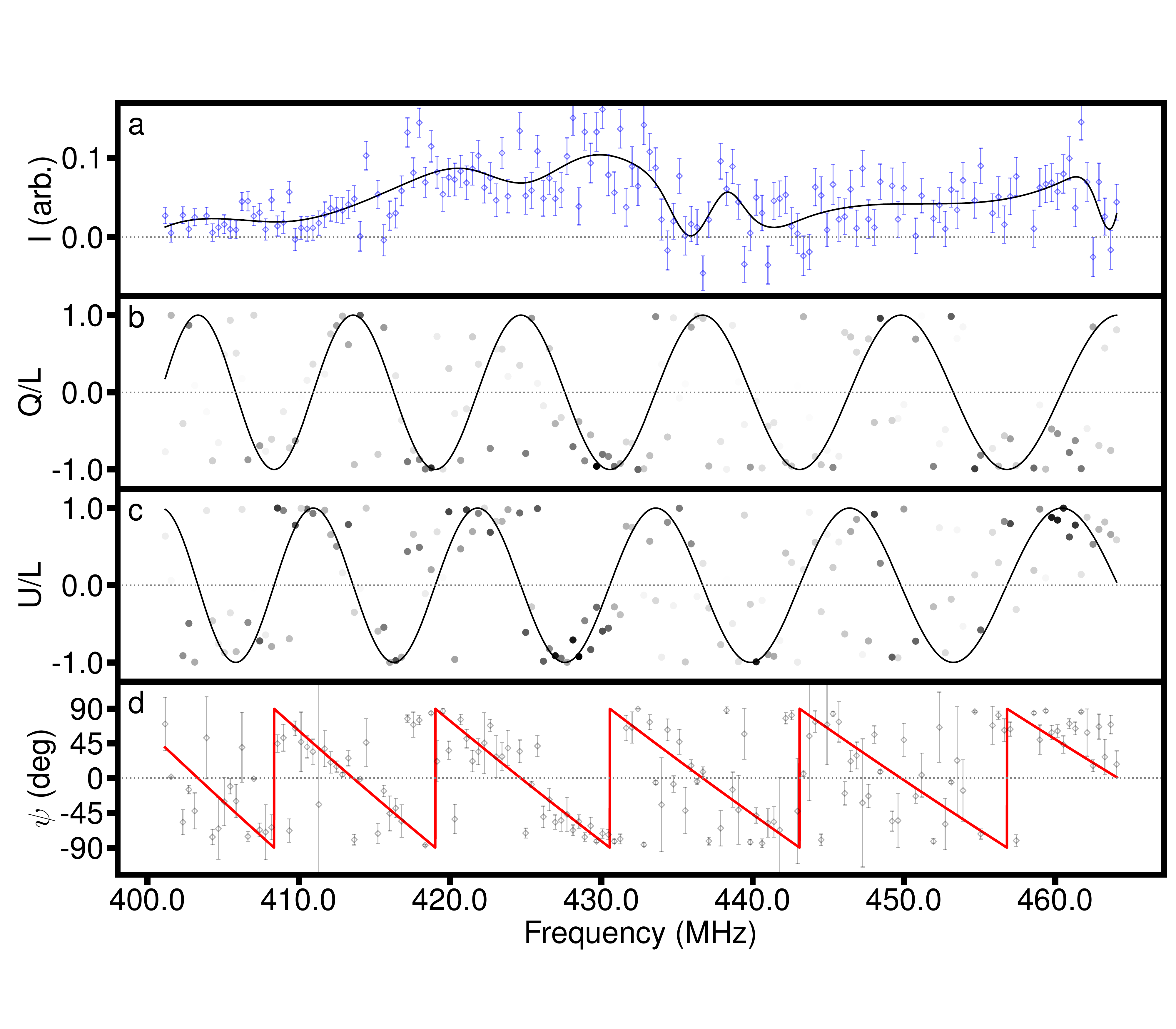}
%
\caption{
\textbf{The polarization spectra for the first observed burst from the ARO 10-m Telescope.}
The spectrum of the first burst in (a) the Stokes I parameter and its cubic spline-smoothed version (black line) (b) the Stokes Q parameter divided by the total linear polarization (L), (c) the Stokes U parameter divided by the total linear polarization, and (d) the uncalibrated polarization angle ($\psi$). The frequency channels with highly polarized signal are indicated with darker points.
The best-fit model of the Faraday rotation modulation with a RM of 116 rad~m$^{-2}$ is indicated with a black line in (b) and (c). The best-fit model of the uncalibrated polarization angle is indicated with the solid (red) line in (d).
}

\label{fig:polarization}
\end{center}
\end{figure}

\end{extended-data}

%% file: acks.tex
\newcommand{\allacks}{
We thank the Dominion Radio Astrophysical Observatory, operated by the National Research Council Canada, for gracious hospitality and useful expertise. The CHIME/FRB Project is funded by a grant from the Canada Foundation for Innovation (CFI) 2015 Innovation Fund (Project 33213), as well as by the Provinces of British Columbia and Qu\'ebec, and by the Dunlap Institute for Astronomy and Astrophysics at the University of Toronto. Additional support was provided by the Canadian Institute for Advanced Research (CIFAR), McGill University and the McGill Space Institute via the Trottier Family Foundation, and the University of British Columbia. 
CHIME is funded by a grant from the CFI Leading Edge Fund (2012)
(Project 31170)  and by contributions from the provinces British
Columbia, Quebec and Ontario. 
The Dunlap Institute is funded by an endowment established by the David Dunlap family and the University of Toronto. Research at Perimeter Institute is supported by the Government of Canada through Industry Canada and by the Province of Ontario through the Ministry of Research \& Innovation. 
The National Radio Astronomy Observatory is a facility of the National Science Foundation operated under cooperative agreement by Associated Universities, Inc.
M.B. is supported by a Fonds de Recherche Nature et Technologie Qu\'ebec (FRQNT)  Doctoral Research Award.
P.C. is supported by an FRQNT Doctoral Research Award.
M.D. is supported by a Killam Fellowship and receives support from an NSERC Discovery Grant, CIFAR, and from the FRQNT Centre de Recherche en Astrophysique du Quebec (CRAQ).
B.M.G. acknowledges the support of NSERC through grant RGPIN-2015-05948, and of the Canada Research Chairs program.
J.W.K. is supported by NSF Award 1458952.
V.M.K. holds the Lorne Trottier Chair in Astrophysics \& Cosmology, a Distinguished James McGill Professorship and receives support from an NSERC Discovery Grant (RGPIN 228738-13) and Gerhard Herzberg Award, from an R. Howard Webster Foundation Fellowship from CIFAR, and from the FRQNT CRAQ.
D.M. is a Banting Postdoctoral Fellow.
S.M.R. is a CIFAR Fellow and is supported by the NSF Physics Frontiers Center award 1430284.
U.-L.P. receives support from Ontario Research Fund—research Excellence Program (ORF-RE), CFI, Simons Foundation, and Alexander von Humboldt Foundation. U.-L.P. acknowledges the support from NSERC, [grant RGPIN-2019-067, CRD 523638-201].
Z.P. is supported by a Schulich Graduate Fellowship from McGill University.
P.S. is a Dunlap Fellow and an NSERC Postdoctoral Fellow.
FRB research at UBC is supported by an NSERC Discovery Grant and by CIFAR.
}